\def\xmas2{X-MAS}
\def\xspec{XSPEC}
\def\mtx{$M$-$T_{\rm X}$} 

\def\thbr{$T_{\rm HBR}$} 
\def\tbroad{$T_{\rm 0.7-7}$} 
\def\thard{$T_{\rm 2.0-7}$}
\def\tx{T_{\rm X}}

\def\ColorRef{[\emph{See the
    electronic edition of the Journal for a color version of this
    figure.}]}
\def\chandra{\emph{Chandra}}
\def\enzo{\emph{Enzo}}

\documentclass[12pt,preprint]{aastex}
\begin{document}
\title{Temperature Structure and Mass-Temperature Scatter In
  Galaxy Clusters}

\author{
  David A. Ventimiglia\altaffilmark{1},
  G. Mark Voit\altaffilmark{1},
  E. Rasia\altaffilmark{2,3}
}


\altaffiltext{1}{Michigan State University, Physics \& Astronomy
  Dept., East Lansing, MI 48824-2320; ventimig@msu.edu,
  voit@pa.msu.edu}
\altaffiltext{2}{Dept. of Astronomy, University of Michigan, Ann
  Arbor, MI 48109, USA}
\altaffiltext{3}{Chandra Fellow}

\begin{abstract}
  Precision cosmology studies based on wide-field surveys of galaxy
  clusters benefit from constraints on intrinsic scatter in
  mass-observable relationships.  In principle, two-parameter models
  combining direct measurements of galaxy cluster structural variation
  with mass proxies such as X-ray luminosity and temperature can be
  used to constrain scatter in the relationship between the mass proxy
  and the cluster's halo mass and to measure the redshift evolution of
  that scatter.  One candidate for quantifying cluster substructure is
  the ICM temperature inhomogeneity inferred from X-ray spectral
  properties, an example of which is \thbr, the ratio of hardband to
  broadband spectral-fit temperatures.  In this paper we test the
  effectiveness of \thbr\ as an indicator of scatter in the
  mass-temperature relation using 118 galaxy clusters simulated with
  radiative cooling and feedback.  We find that, while \thbr\ is
  correlated with clusters' departures $\delta \ln T_X$ from the mean
  \mtx\ relation, the effect is modest.
\end{abstract}


\section{Introduction}

Galaxy clusters play an important role in precision cosmology that
complements other techniques like Type Ia supernovae
luminosity-distance relation measurements
\citep{perlmutter_measurements_1999,riess_observational_1998},
baryonic acoustic oscillations (BAOs) angular-distance relation
measurements \citep{eisenstein_detection_2005}, and observations of
the cosmic microwave background radiation (see
\cite{frieman_dark_2008} for a review).  For example,
\cite{vikhlinin_chandra_2009} exploit dark energy's influence on the
growth of structure by using 37 moderate-redshift and 49 low-redshift
clusters to measure the shape of the galaxy-cluster mass function and
its redshift evolution, which constrain the dark-energy density
parameter $\Omega_{\Lambda}$ to $0.83 \pm 0.15$ in a non-flat
$\Lambda\rm{CDM}$ cosmology and the dark energy equation of state
parameter $w_0$ to $-1.14 \pm 0.21$ in a flat cosmology.
\cite{2010MNRAS.406.1759M} have obtained similar results from
measurements of the evolving number density of the largest clusters in
order to constrain $w_0$ to $-1.01 \pm 0.20$.

Strategies like these that compare model predictions to galaxy-cluster
sample statistics inevitably confront sample error and sample bias.
Future surveys expected to gather samples of 10--40 thousand galaxy
clusters\footnote{http://www.mpe.mpg.de/heg/www/Projects/EROSITA/main.html}
will maximize survey reach while maintaining sufficient observation
quality in order to minimize sample error, but they still must grapple
with a major source of sample bias, which is scatter in the
relationship used to infer cluster mass from an observable mass proxy.
An important mass-observable relation for galaxy cluster studies
connects dark matter halo mass to the temperature of the intracluster
medium (ICM) inferred from its X-ray spectrum (its ``X-ray
temperature'' $T_X$).  In this paper we investigate the possibility of
correcting for scatter in this relation using
temperature-inhomogeneity in the ICM, and discuss challenges that may
exist in such a program.

A significant amount of uncertainty in the dark-energy constraints
obtainable from large cluster surveys derives from uncertainty in
scatter about the mean scaling relations obeyed by galaxy clusters'
bulk properties
\citep{lima_self-calibration_2005,2010PhRvD..81h3509C}.  The key
galaxy cluster property to measure when trying to constrain dark
energy with clusters is the cluster's mass, which cannot be directly
observed.  Theoretical considerations predict correlations among halo
mass and more readily observed cluster properties, like its galaxy
richness, the velocity dispersion of its galaxies, $T_X$, the
Sunyaev-Zel'dovich decrement, the gas mass, and $Y_{\rm X}$ parameter,
which is the product of $T_X$ and the gas mass inferred from X-ray
observations \citep{kravtsov_new_2006}.  Theory also predicts
intrinsic scatter in these relations owing to variation in cluster
dynamical state \citep[see, for example][]{stanek.etal.10,fabjan.etal.11,rasia.etal.11}.

One way to deal with intrinsic scatter is to join a cluster model to a
cosmological model and simultaneously fit for the parameters of both, 
leveraging the statistical power of large surveys and ``self-calibrating'' the
mass-observable relations.  Another approach is to combine observables
that tend to depart from the expected scaling relations in opposite
ways, yielding a new, low-scatter composite observable.  An example
low-scatter composite observable is $Y_{\rm X}$, since at a given halo
mass, offsets in the measured gas mass at fixed total mass tend to
anti-correlate with offsets in the measured temperature
\citep{kravtsov_new_2006}.

Another family of low-scatter composite observables attempt to measure
structural variation directly.  Mergers, relaxation, and non-adiabatic
processes like radiative cooling, star formation, and feedback ought
to leave a visible imprint that may allow us to measure and correct
for scatter.  For example, one might use imaging to quantify resolved
morphological substructure.  \cite{jeltema_cluster_2008} apply two
observationally-motivated structure measures, the power ratios
\citep{buote_quantifyingmorphologies_1995} and the centroid shift
\citep{mohr_x-ray_1993}, to a sample of galaxy clusters simulated with
\enzo\ \citep{norman_cosmological_1999,oshea_introducing_2004}.  They
find that cluster structure correlates strongly with bias in mass
estimates derived from $T_X$ under the assumption of hydrostatic
equilibrium and accounting for cluster structure can be used to
correct some of the bias.  Similarly,
\cite{ventimiglia_substructure_2008} apply the power ratios, centroid
shift, and axial ratio \citep{ohara_effects_2006} substructure
measures to the same sample of simulated clusters used in this paper,
and find that cluster substructure correlates with departures from the
mean \mtx\ relationship in the sense that clusters with more
  substructure tend to have a lower temperature at a given mass, and
can be used to refine mass estimates derived from the ICM X-ray
temperature.  \cite{piffaretti_valdarnini_substructure_2008}
  likewise find that greater substructure, as quantified with power
  ratios, correlates with lower temperature at a given mass. 
\cite{yang_influence_2009} find a strong correlation between
mass-temperature scatter and halo concentration in their sample of
simulated clusters, with cooler clusters appearing more concentrated
than warmer clusters at similar mass.

\cite{jeltema_cluster_2008} observe, however, that line-of-sight
projection effects lead to significant uncertainties in
morphologically-derived substructure measures.  Substructure also
becomes more difficult to resolve at high redshift.  Spectral
signatures of dynamical state are therefore attractive because they
are aspect-independent and redshift-independent.  One such spectral
signature of dynamical state is the ``temperature ratio'' \thbr\
\citep{mathiesen_four_2001,cavagnolo_bandpass_2008}, which divides a
``hardband'' spectral-fit temperature by a ``broadband'' spectral-fit
temperature.  An energy cut applied to a broadband spectrum produces a
hardband spectrum and serves to filter out cooler line-emitting
components of the ICM.  \cite{mathiesen_four_2001} studied the effects
of relaxation on the observable properties of galaxy clusters, using a
sample of numerically-simulated galaxy clusters generated by
\cite{mohr_x-ray_1997}.  They found that hardband (2.0--9.0 keV) X-ray
spectral fit temperatures average $\sim$20\% higher than broadband
(0.5--9.0 keV) temperatures and suggested that this effect may signal
the presence of cool, luminous sub-clusters lowering the broadband
temperature.  
\cite{valdarnini_substructure_2006} corroborated these findings in simulations that included radiative cooling.

\begin{figure}[p]
  \centering
  \includegraphics[width=0.7\textwidth,angle=-90]{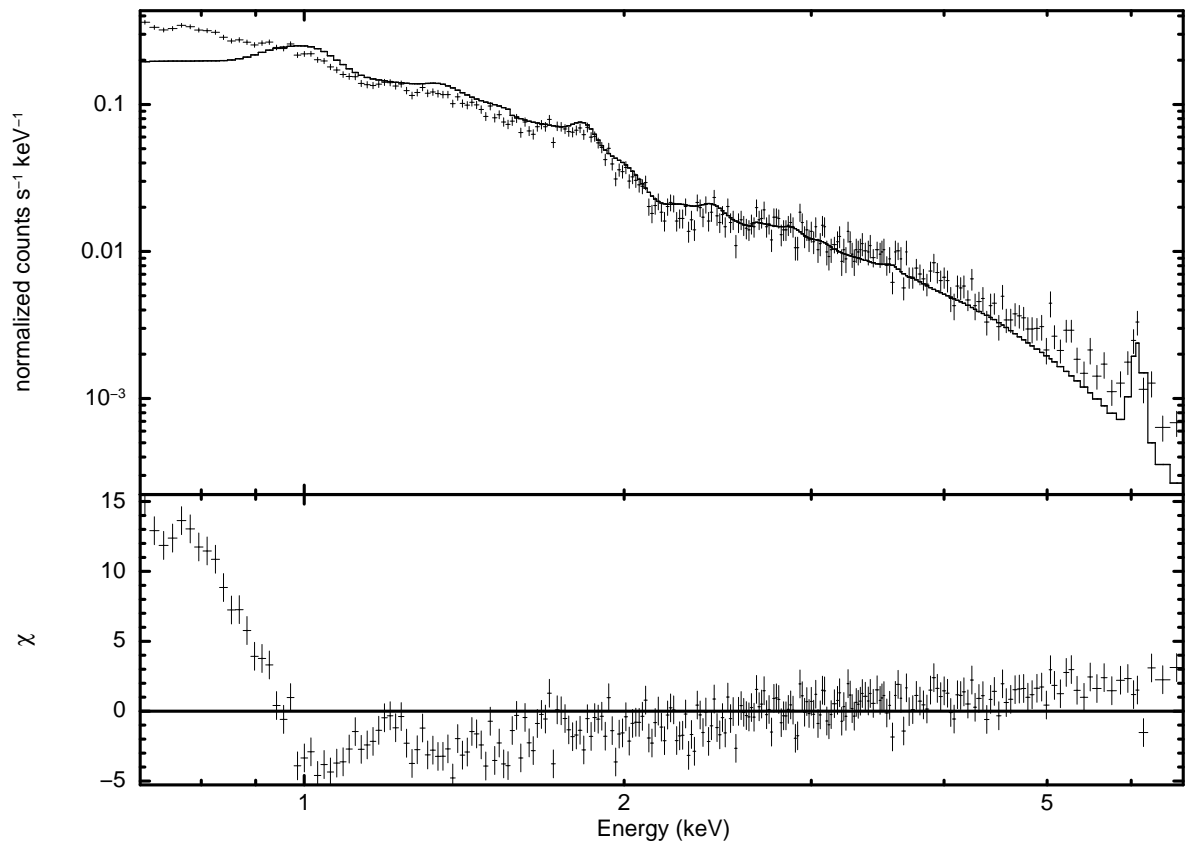}
  \caption{\xmas2 simulated counts spectrum for a simulated cluster
    that appears to have significant temperature structure.  A
    single-temperature MEKAL model fit to the [0.7--7.0] keV
    broad band is over-plotted as the solid line, with
    $k_{B}T=2.42\pm0.03$ keV.  Fit residuals appear in the bottom
    panel.  This figure and Figure \ref{fig:f02} illustrate
    qualitatively the effect that additional cool components have on a
    single-temperature fit.}
  \label{fig:f01}
\end{figure}

\begin{figure}[p]
  \centering
  \includegraphics[width=0.7\textwidth,angle=-90]{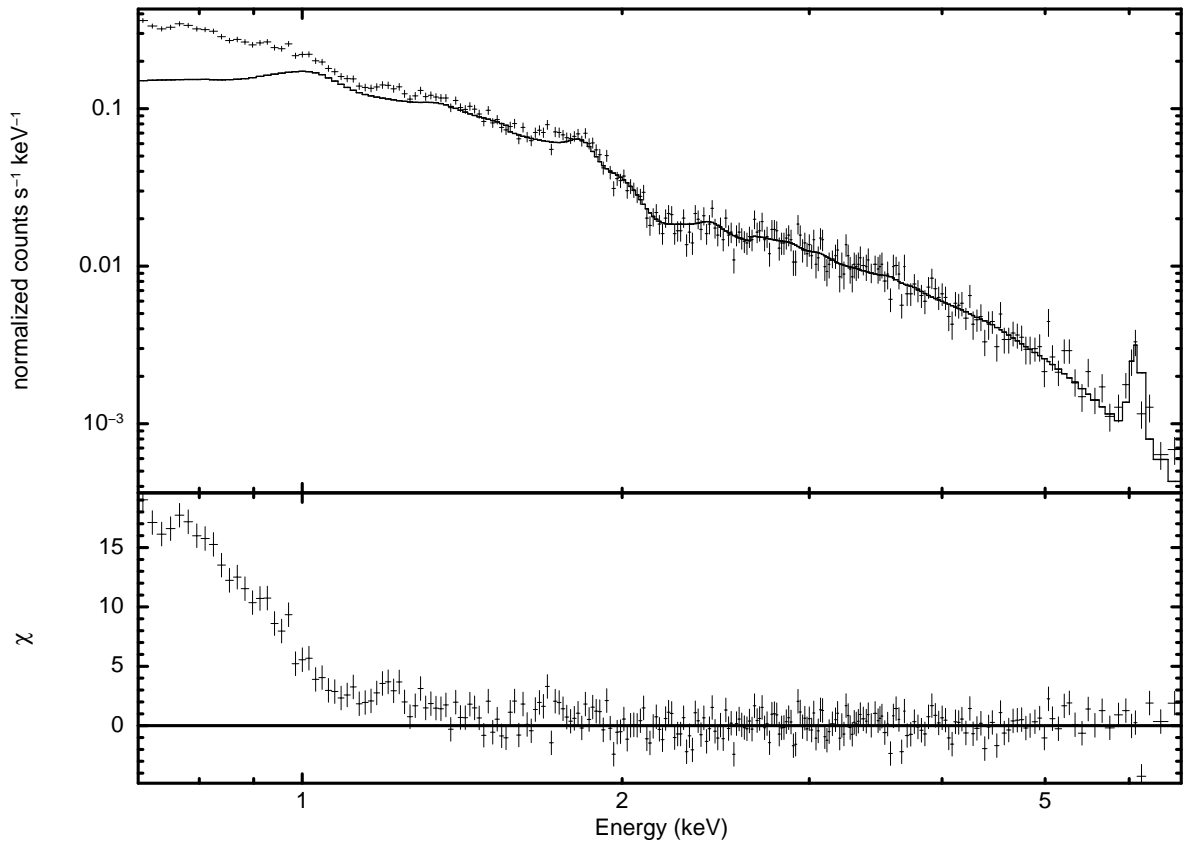}
  \caption{\xmas2 simulated counts spectrum for the same simulated
    cluster as in the previous figure.  A single-temperature MEKAL
    model fit to the [2.0--7.0] keV hard band is over-plotted as the
    solid line, with $k_BT=3.52\pm0.2$ keV.  Fit residuals appear in
    the bottom panel.}
  \label{fig:f02}
\end{figure}

Figures \ref{fig:f01} and \ref{fig:f02} illustrate the
effect.  Both show a counts spectrum and single-temperature fit for a
typical, unrelaxed cluster in our sample, with an aperture set at
$R_{2500}$ and a core region out to $0.15 R_{2500}$ excised.  Figure
\ref{fig:f01} is for a single-temperature model fit to the broad band,
while figure \ref{fig:f02} is for the hard band.  Note the excess
emission relative to the model above 4.0 keV and the deficit below 2.5
keV that arises because the model cannot simultaneously fit both a hot
component and a cooler, line-emitting component
\citep{2004MNRAS.354...10M}. Figure \ref{fig:f02} shows the same
spectrum, but with the residuals for a model fit just to the hard
band.  In this case the fit is much better over the range from 2.0 to
7.0 keV but under-predicts the emission below 2.0~keV.

\cite{mathiesen_four_2001} suggested that the temperature skewing they
observed might indicate a real temperature skewing detectable in real
clusters using \chandra.  \cite{cavagnolo_bandpass_2008} fit
single-temperature emission models to the hard band (2.0--7.0 keV) and
broad band (0.7--7.0 keV) for a large (N = 192) sample of clusters with
observations selected from the \chandra\ Data Archive.  These authors
show that for a large, heterogeneous sample of clusters across a broad
temperature range, the distribution of \thbr\ has a mean of 1.16 and
an rms deviation $\sigma = \pm0.10$, with \thbr\ tending to be larger
in merging systems.  They also report that while this signal is
significant in the aggregate, the errors for any single \thbr\
measurement and the scatter across all of the measurements together
pose challenges for any effort to use \thbr\ either to select for
merging systems or to obtain more accurate mass estimates.  In order
to meet these challenges it is important also to examine \thbr\ and
similar spectral signatures of dynamical state in a simulation
context.  Like the original study of \cite{mathiesen_four_2001}, this
paper examines the temperature ratio for a sample of simulated
clusters and simulated \chandra\ observations, 
and as in the subsequent study by \cite{valdarnini_substructure_2006}, 
the simulated clusters analyzed here were generated using a hydrodynamical code with radiative cooling included.

This paper is organized as follows.  In $\S2$, we describe the
simulated clusters in our sample along with the \xmas2 code for
generating their mock X-ray observations.  In $\S3$, we present our
analysis methods and define the spectral signatures of dynamical state
we examine.  In $\S4$, we report and discuss our results, and $\S5$
summarizes our work.

\section{Methods}

\subsection{Numerical Simulations}
This study is based on an analysis of 118 clusters simulated using the
cosmological hydrodynamics TREE+SPH code GADGET-2
\citep{springel_cosmological_2005}, which were simulated in a standard
$\Lambda$ cold dark matter ($\Lambda$CDM) universe with matter density
$\Omega_M$ = 0.3, $h$ = 0.7, $\Omega_b$ = 0.04, and $\sigma_8$ = 0.8.
The simulation includes radiative cooling assuming an optically-thin
gas of primordial composition, with a time-dependent UV background
from a population of quasars, and handles star formation and supernova
feedback using a two-phase fluid model with cold star-forming clouds
embedded in a hot medium.  All but four of the clusters are from the
simulation described in \cite{borgani_hydrodynamical_2004}, who
simulated a box $192 \, h^{-1} \, {\rm Mpc}$ on a side, with $480^3$
dark matter particles and an equal number of gas particles.  The
present analysis considers the 114 most massive clusters within this
box at $z = 0$, which all have $M_{200}$ greater than $5 \times
10^{13} \, h^{-1} M_\odot$.  These are referred to as the B04 Sample
in the remainder of this paper.  By convention, $M_{\Delta}$ refers to
the mass contained in a sphere which has a mean density of $\Delta$
times the critical density $\rho_c$, and whose radius is denoted by
$R_{\Delta}$

That cluster set covers the $\sim$1.5-5 keV temperature range, but the
$192 \, h^{-1} \, {\rm Mpc}$ box is too small to contain significantly
hotter clusters.  We therefore supplemented it with four clusters with
masses $> 10^{15} \, h^{-1} \, M_\odot$ and temperatures $> 5$ keV
drawn from a dark-matter-only simulation in a larger $479 \, h^{-1} \,
{\rm Mpc}$ box \citep{2009MNRAS.398.1678D}, referred to in this paper
as the D09 sample.  The cosmology for this simulation also was
$\Lambda$CDM, but with $\sigma_8$ = 0.9.  These were then re-simulated
including hydrodynamics, radiative cooling, and star formation, again
with GADGET-2 and using the zoomed-initial-conditions technique of
\cite{tormen_rise_1997}, with a fourfold increase in resolution.  This
is comparable to the resolution of the clusters in the smaller box.
Adding these four massive clusters to our sample gives a total of 118
clusters with $M_{200}$ in the range $5 \times 10^{13} \, h^{-1}
M_\odot$ to $2 \times 10^{15} \, h^{-1} M_\odot$.
The mean structural properties of massive clusters drawn from a sample with $\sigma_8$ = 0.9 may differ somewhat from those of similar-mass clusters in a $\sigma_8$ = 0.8 universe because they reflect a more advanced state of cosmic evolution.  However, these four additional clusters carry minimal statistical weight in the context of the overall sample.  They are included primarily to evaluate whether the mean and dispersion of their $T_{\rm HBR}$ values are consistent with those of the lower-mass systems.

\subsection{X-Ray Simulations}

The simulated galaxy clusters in our sample are processed with the
\emph{X-ray Map Simulator} version 2 (\xmas2)
\citep{gardini_simulating_2004,rasia_x-mas2:_2008} to generate X-ray
images suitable for standard \chandra\ reduction techniques.  In its
first step \xmas2 uses the outputs of the hydrodynamic code to
calculate the emissivity of each simulation element within the chosen
field of view and to project this onto the image plane.  In its second
step it convolves the resulting flux with the appropriate response of
a given detector.  In the case of our simulated \chandra\
observations, the second step applies the response matrix file and
ancillary response file for the ACIS S3 CCD, with a 200 kilosecond
exposure time.  In order to separate out potential calibration issues
from the focus of this particular study, these response matrices
implement a constant response over the detector, and in generating the
simulated X-ray images, the evolved zero-redshift clusters are shifted
to a redshift sufficient to fit $R_{500}$ within the 16 arcminute
field of view.  Because of this step, though the clusters in the
sample are distributed over a range of physical sizes, they all have
approximately the same apparent size projected onto the image plane
and into the simulated observations.

\subsection{\mtx\ Relation}
In this paper, cluster mass refers to $M_{200}$, while $T_X$ refers to
our estimated ``spectral-fit temperatures'' obtained using \xspec\
\citep{1996ASPC..101...17A} to fit single-temperature MEKAL plasma
models to various energy bands in simulated \xmas2 X-ray observations
taken from the numerically-simulated clusters.  The steps involved in
the process are described in detail in $\S$ \ref{sec:analysis}.

\begin{figure}[p]
  \centering
  \includegraphics[width=0.8\textwidth, trim=0mm 0mm 0mm 0mm, clip=true]{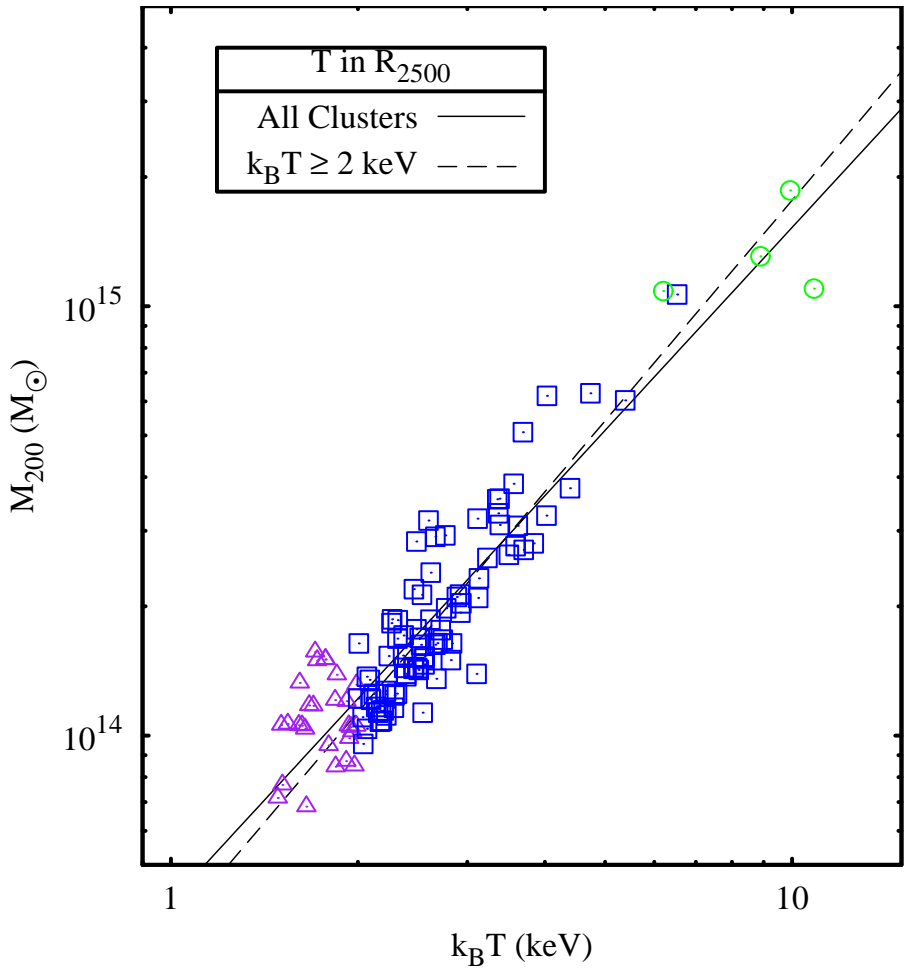}
  \caption{Mass-temperature (\mtx) relation for the clusters in our
    sample.  Clusters with $k_BT <$ 2 keV are plotted in purple open
    triangles, while clusters with $k_BT \ge$ 2 keV are plotted in
    blue open squares, and the four massive D09 clusters in green open
    circles.  Mean relations are plotted with solid lines for the
    whole sample, and dashed lines for clusters with $k_BT \ge$ 2keV.
    Finally, note that average temperatures are taken within a
    core-excised annulus whose outer radius is placed at $R_{2500}$.
    \ColorRef}
  \label{fig:f03}
\end{figure}

Figure \ref{fig:f03} shows the mass-temperature relation based on our
sample of simulated clusters and estimated spectral fit temperatures,
measured in a broad 0.7--7 keV band within an annular aperture from
$0.15 R_{500}$ to $R_{2500}$ as in \cite{cavagnolo_bandpass_2008}.
The best fits to the power-law form
\begin{equation}
  M = M_0 \left( \frac {\tx} {2 \, {\rm keV}} \right)^\alpha
  \label{eq:mass-temp}
\end{equation}
have the coefficients $M_0 = 1.09\pm0.01 \times 10^{14}\ h^{-1}
M_{\odot}$, $\alpha = 1.57\pm0.06$ for the full sample of clusters.
For the subset of clusters with \thard$>2.0$kev, the best-fit parameters 
are $M_0 = 1.06\pm0.02 \times 10^{14}\
h^{-1} M_{\odot}$, $\alpha = 1.71\pm0.07$.  As is generally the case
for simulated clusters, the power-law indices of the mass-temperature
relations found here are consistent with cluster self-similarity and
the virial theorem
\citep{kaiser_evolution_1986,navarro_simulations_1995}.  These
relationships have scatter, which we characterize by the standard
deviation in log space $\sigma_{\ln M}$ about the best-fit mass at
fixed temperature $\tx$.  We find $\sigma_{\ln M} = 0.10$.

\section{Analysis}
\label{sec:analysis}

\subsection{Filtering}
\label{subsec:specextract}

We use \xmas2 to simulate X-ray observations---provided as standard
\chandra\ event files---then subject them to a series of reduction
steps using the \chandra\ Interactive Analysis of Observations package
(CIAO) v4.1 \citep{fruscione_ciao:_2006}.  All of the event files have
at least 100K counts, while those for the most luminous clusters have
over half a million counts.  Simulated observations of this quality
provide us with an opportunity to address the question of how much of
the scatter in our temperature-structure statistics is intrinsic.
Consequently, our first reduction step filters each raw event file
into a new set of event files by randomly sampling the events it
records.  The number of counts in each file is its ``count level,''
and in order to cover the space of typical \chandra\ archival
observations, we apply the filtering step to each of the raw event
files four times, at count levels 15K, 30K, 60K, and 120K.  These
count levels roughly map to the range between observations of short
duration or of relatively low surface-brightness objects to
observations of long duration or of relatively high surface-brightness
objects.

\subsection{Cool Lump Excision}
\label{sec:cool_lump_intro}

\begin{figure}[p]
  \centering
  \includegraphics[width=1.0\textwidth, trim=0mm 0mm 0mm 0mm, clip=true]{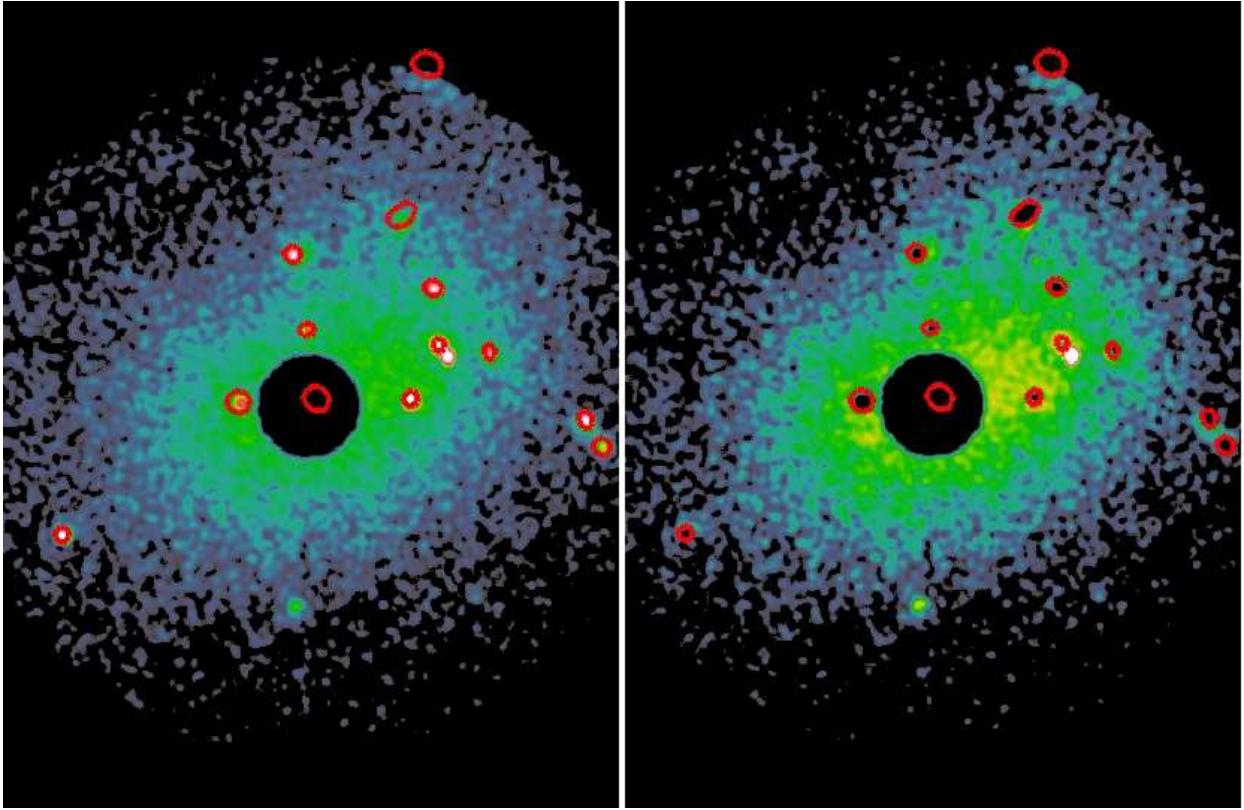}
  \caption{\xmas2 simulated X-ray surface-brightness images for the
    same cluster whose spectrum is presented in Figures \ref{fig:f01}
    and \ref{fig:f02}.  The aperture is set at $R_{2500}$ and a core
    region of size $0.15R_{2500}$ is excised.  Note the
    surface-brightness peaks in the image.  In this study these are
    detected automatically by the CIAO tool wavdetect, which generates
    source regions identified in this figure by red ellipses.  Labeled
    ``cool lumps'' in this analysis, they are subjected to varying
    degrees of excision, from no masking (\emph{left}) to full masking
    (\emph{right}), as described in the text.  \ColorRef}
  \label{fig:f04}
\end{figure}

An example of a filtered event file appears in Figure \ref{fig:f04} as
a surface-brightness image.  It displays a common feature of numerical
simulations of galaxy clusters, which is the presence of relatively
dense, cool, and metal-rich substructures that continuously undergo
mass accretion and have not yet come into thermal equilibrium with the
hot ICM surrounding them \citep{2009arXiv0906.4370B}.  These bright
point-like spots, or ``cool lumps'', typically are associated with the
dense cool cores of smaller halos that have merged with the primary
halo.  Generally regarded as an unphysical artifact of numerical
simulations, at least insofar as their temperatures, densities, and
concentrations are concerned, commonly they are excised before further
analysis is conducted \citep{2006MNRAS.369.2013R, piffaretti_valdarnini_substructure_2008, nagai_effects_2007}.

In order to study the effect that excising cool lumps has on measures
of temperature substructure, we produce from the originals several new
event files whose cool lumps have been excised to an increasing
degree.  The CIAO \verb|wavdetect| tool identifies peaks in the photon
distribution by correlating an event file's image with a sequence of
``Mexican-Hat'' wavelet functions of differing scale sizes, measured
in pixels, then generates a source list with associated region files.
We use its default sequence of wavelet scale sizes in this analysis (2
and 4 pixels), though we apply the tool four times to each filtered
event file, each time incrementing the multiplicative factor by which
the source regions are scaled.  In CIAO \verb|wavdetect|, the
parameter governing this multiplicative factor is \verb|ellsigma| and
in our study ranges between 0 to 3. A value of 0 for \verb|ellsigma|
is equivalent to ``no masking'' while a value of 3 corresponds to what
we call ``full masking''.  The effect is to produce versions of each
cluster observation with a range of masking. 
Note that the wavelet scale size is a constant fraction of the cluster size because all the clusters have been redshifted so that $R_{500}$ fits in the field of view.

We finish the extraction phase of our analysis with the following
steps.  First, for every excised file we apply an aperture of
$R_{2500}$ and excise the central $0.15\ R_{2500}$ region, generating
new copies of the event files.  This step, including the centering algorithm,
matches the procedure in
\citet{cavagnolo_bandpass_2008} so that regions measured in the
simulated clusters correspond to those in the \chandra\ archival
observations.  Next, each fully-processed event file file is extracted
into a standard pulse-invariant (PI) spectral file binned so as to
have at least 25 counts in each energy channel.  The extraction phase
is complete when each original raw event file generates PI spectral
files suitable for spectral fitting in \xspec.

\subsection{Spectral Fitting}

The PI spectral files are then fed into \xspec\ v12.5.0 for spectral
fitting.  In order to form the temperature ratio \thbr\ two fits are
performed for each spectrum.  The first fit is over the hard band from
2.0$(1+z)^{-1}$--7 keV, for which the $(1+z)^{-1}$ factor exists in
order to shift the 2 keV cutoff from the observer's frame to the
cluster's rest-frame. The simulated clusters occupy a range of
redshifts because larger clusters are translated to higher redshifts
in order to fit $R_{500}$ within 16 arc-minutes, when creating
artificial observations.  The second fit is over the broad band,
including all energy channels in the spectrum from 0.7--7 keV.  Every
fit is made by minimizing the $\chi^2$ statistic for a
single-temperature MEKAL model multiplied by a warm-absorber (WABS, to
account for Galactic absorption).  The Galactic column is fixed to
$N_H = 5 \times 10^{20}\ {\rm cm}^{-2}$ and the metallicity to 0.3
Solar, leaving the temperature of the emission component, its H
density (although this makes no difference), and its normalization as
the only free parameters.

\subsection{Quantifying Temperature Structure}
\label{sec:coolresid}

We adopt two spectral measurements of temperature structure.  The
first is the temperature ratio \thbr\ of
\cite{cavagnolo_bandpass_2008} and is found in the following way.  For
each simulated cluster in our sample, for each count level (15K, 30K,
60K, 120K), and for each cool lump masking level (0, 1, 2, 3), we find
a spectral-fit temperature in the hard band and the broad band and
form the temperature ratio \thbr.  Emission from cooler, line-emitting
metal-rich parts of the ICM ought to be excluded from the hard band,
so we expect that a \thbr\ value greater than unity signals the
presence of merging sub-clumps.

Our second measure of temperature, which we call the ``cool
residual,'' ($RES_{cool}$) compares an observation's actual broadband
count rate to a model-predicted count rate for a spectral model fit
only over the hard band.  Again, we determined a spectral-fit
temperature for each combination of cluster, count level, and masking
level, except that in this case we performed only a hardband fit.
Since emission from cooler components should be excluded from this
band, in general we achieve good fits even when the count level is
sufficiently high that broadband fits may formally have large reduced
$\chi^2$ values.  From this hardband model fit we estimate the
corresponding broadband count rate and calculate its percentage
deviation from the actual broadband count rate.  While
single-temperature systems will have actual count rates that are
essentially the same as their model count rates, the introduction of
cooler, luminous substructure components should generate an excess
broadband count rate relative to the model.  This measure of
temperature substructure avoids the uncertainties associated with a
single-temperature fit to the broad band.

\section{Results and Discussion}
\label{sec:results}

Having calculated spectral-fit temperatures, \thbr, and the cool
residual for all of the clusters in our sample, we have the means
to study how well temperature structure correlates with departures
from the mean \mtx\ relation, and how well these measures can be used
to obtain better mass estimates.  We will also examine how our results
depend on the removal of the cool lumps.

\subsection{\thbr\ from Simulations.}

\begin{figure*}[p]
  \centering
  \includegraphics[width=0.9\textwidth, trim=0mm 30mm 0mm 30mm, clip=true]{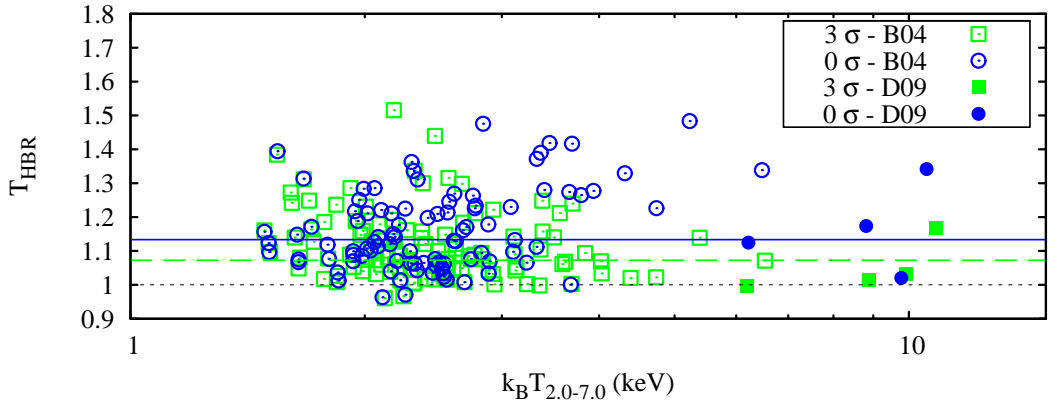}
  \caption{\thbr\ plotted against \tbroad\ for simulated \xmas2
    observations that have approximately 120K counts.  Hydrodynamic
    simulations may produce spurious over-condensations of cool gas.
    These cool lumps can be excised using the CIAO tool wavdetect,
    whose aggressiveness can be controlled via its
    ellsigma parameter.  Squares correspond to
    single-temperature MEKAL fits whose underlying observations are
    processed by wavdetect with ellsigma set to 3,
    and circles correspond to those with ellsigma
    set to 0. Finally, note that the 4 massive D09 clusters are
    denoted by filled symbols rather than by open symbols.  The solid
    line represents the mean when cool lumps are not removed, while
    the dashed line represents the mean when they are removed.  The
    dotted line indicates \thbr\ = 1.  \ColorRef}
  \label{fig:f05}
\end{figure*}

We present in Figure \ref{fig:f05} the temperature ratios \thbr\ for
our simulation sample, plotted as a function of the broadband
temperature fit \tbroad.  This figure is similar to Figure 8 in
\cite{cavagnolo_bandpass_2008}, which shows that the mean value
of \thbr\ observed among clusters in the {\em Chandra} archive is
$\langle T_{\rm HBR} \rangle = 1.16$, with a standard deviation of
$\sigma_{\rm THBR} = 0.10$.  We remind the reader of important
differences between the two samples being considered.  The \chandra\
archive sample of \cite{cavagnolo_bandpass_2008} contains clusters
most of which have $k_BT >$ 3 keV, whereas most of the clusters in our
sample have $k_BT <$ 3 keV.  Nevertheless, even with that caveat the
temperature ratios of the simulated clusters are distributed in
approximately the same way as are those in the real sample of
\cite{cavagnolo_bandpass_2008}, with $\langle T_{\rm HBR} \rangle =
1.12$ and $\sigma_{\rm THBR} = 0.11$, provided the cool lumps are not
excised (circles in Figure \ref{fig:f05}).  When the cool lumps are
excised (squares in Figure \ref{fig:f05}), the mean and variance of
\thbr\ diminish, with $\langle T_{\rm HBR} \rangle = 1.07$ and
$\sigma_{\rm THBR} = 0.07$.  These values are for the full 120K count
data, though the full set of $\langle$\thbr$\rangle$ for the four
\verb|ellsigma| values and four count values are tabulated in Table
\ref{tbl:01}.  This table provides mean values for \thbr\ and its
variance for all of the clusters in the sample, and for a subset whose
\thard\ value is greater than 2 keV.  Much of the variance derives
from the lower-temperature clusters, and as they are removed the
variance in \thbr\ drops significantly, especially when full masking
is applied.

\begin{figure*}[p]
  \centering
  \includegraphics[width=0.9\textwidth, trim=0mm 30mm 0mm 30mm, clip=true]{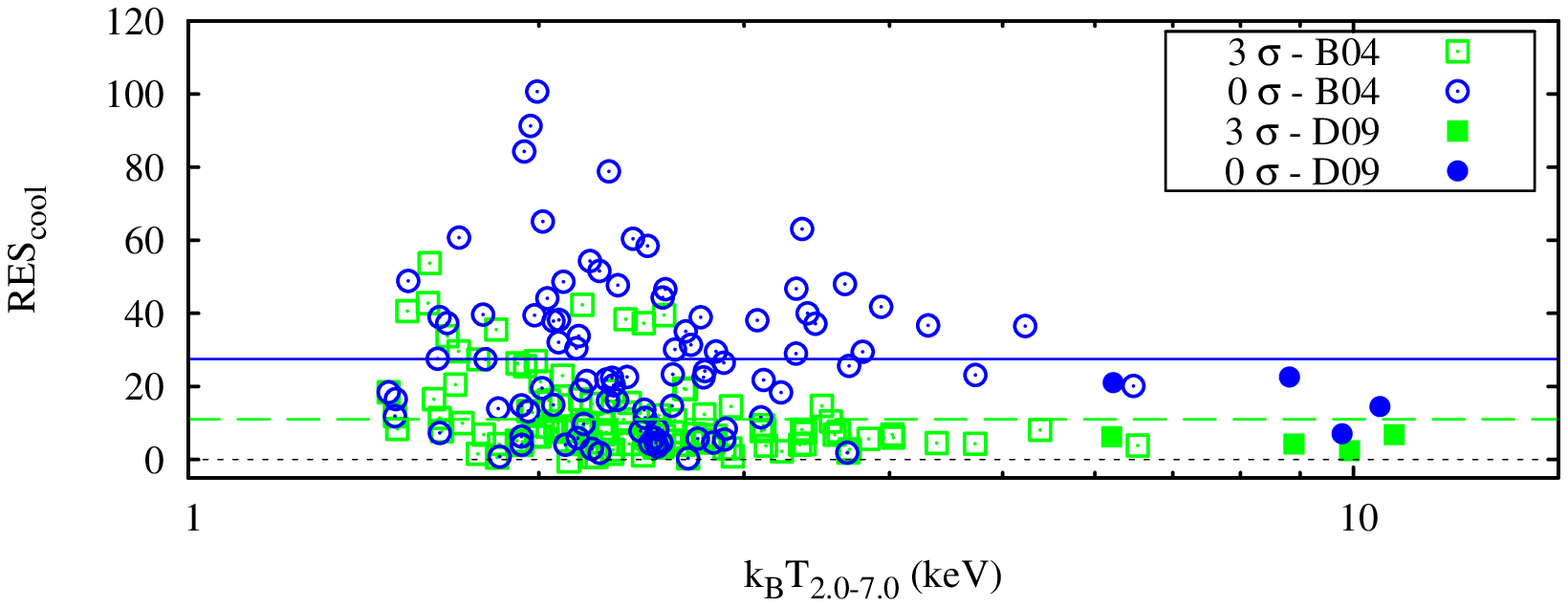}
  \caption{$RES_{cool}$ plotted against \tbroad\ for simulated \xmas2
    observations that have approximately 120K counts.  This statistic
    tracks the percent excess count rate of the actual observation
    with respect to count rate for a MEKAL model fit to just the hard
    band.  Squares correspond to single-temperature MEKAL fits whose
    underlying observations are processed by wavdetect with
    ellsigma set to 3, and circles correspond to
    those with ellsigma set to 0. As in the previous
    plot, the 4 massive D09 clusters are denoted by filled symbols
    rather than by open symbols.  The solid line represents the mean
    when cool lumps are not removed, the dashed line represents the
    mean when they are removed, and the dotted line indicates zero
    residual.  \ColorRef}
  \label{fig:f06}
\end{figure*}

Figure \ref{fig:f06} presents a similar effect for $RES_{cool}$.
Here, we see that the excision of cool lumps again reduces the mean
and variance of the temperature substructure measure, although in this
case the effect is more dramatic.  Evidently, the presumably
unphysical cool lumps in simulated clusters may be necessary to reproduce
the distribution of quantitative temperature substructure measures
found in real clusters.  This is a subject which we return to in
Section \ref{sec:masking}.

\subsection{Temperature Structure and Scaling Relations}
\label{sec:mtx}

\begin{figure}[p]
  \centering
  \includegraphics[width=0.8\textwidth, trim=0mm 0mm 0mm 0mm, clip=true]{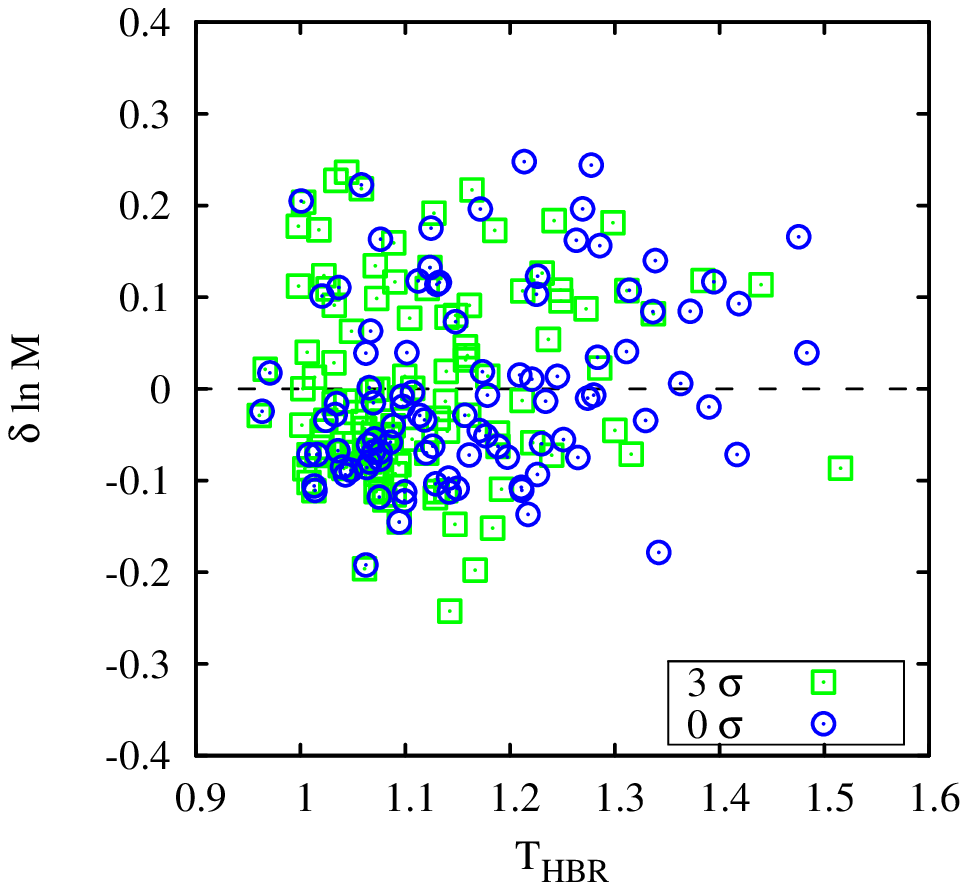}
  \caption{Relationship between \thbr\ and mass offset $\delta\ln M
    (\tx)$ from the mean \mtx\ relationship for the clusters in
    combined B04+D09 simulation sample.  The temperatures in this
    relation are our broadband spectral-fit temperatures.  Squares
    correspond to simulated \xmas2 observations that are processed by
    wavdetect with ellsigma set to 3 (full masking), while circles
    correspond to observations that have ellsigma set to 0.
    \ColorRef}
  \label{fig:f07}
\end{figure}

\begin{figure}[p]
  \centering
  \includegraphics[width=0.8\textwidth, trim=0mm 0mm 0mm 0mm, clip=true]{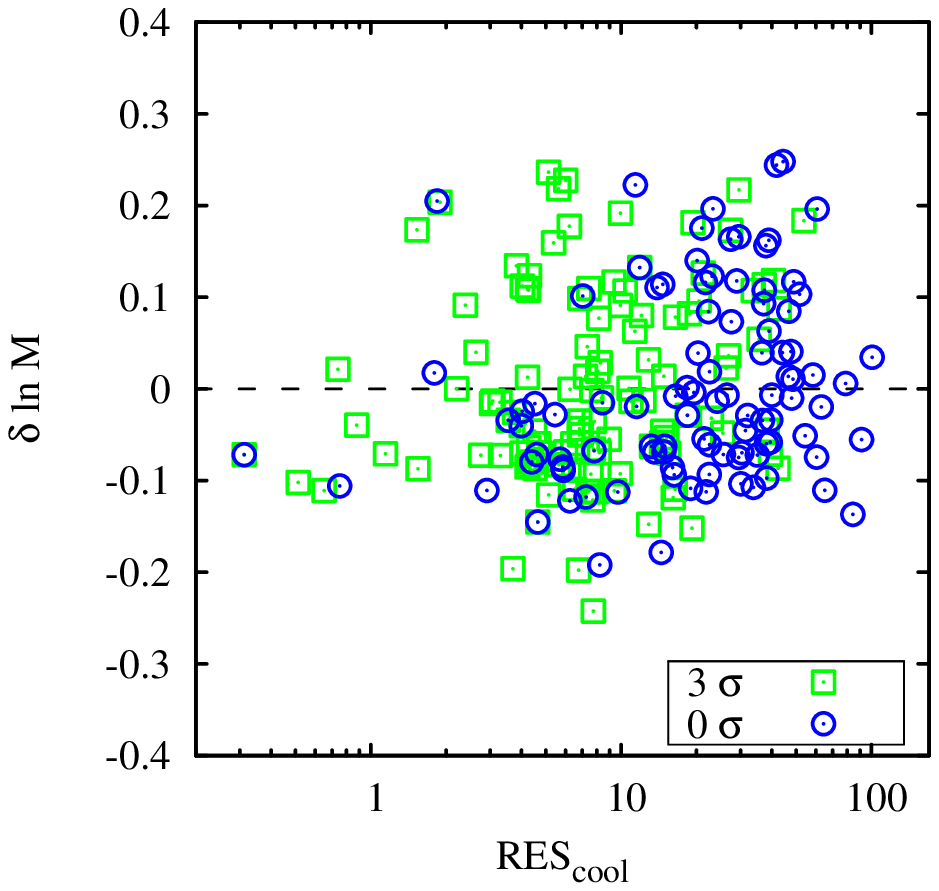}
  \caption{Relationship between $RES_{cool}$ and mass offset
    $\delta\ln M (\tx)$ from the mean \mtx\ relationship for the
    clusters in combined B04+D09 simulation sample.  The temperatures
    in this relation are our broadband spectral-fit temperatures.
    Squares correspond to simulated \xmas2 observations that are
    processed by wavdetect with ellsigma set to 3, while circles
    correspond to observations that have ellsigma set to 0.
    \ColorRef}
  \label{fig:f08}
\end{figure}

Our original motivation for conducting this study was to determine if
temperature structure, as quantified by the temperature ratio \thbr,
correlates with and can be used to correct for departures from the
mean mass-temperature relation \mtx.  In order to test this idea, we
define the ``mass offset'' at fixed temperature by the relation
\begin{equation}
  \delta \ln M (\tx) = \ln \left[ \frac {M} {M_{\rm pred}(\tx)} \right]
  \label{eq:deltaM}
\end{equation}
where $M$ is the cluster's actual mass, and $M_{\rm pred}$ is the mass
predicted from the mean \mtx\ relation.  Figure \ref{fig:f07} plots
the mass offsets for our simulated clusters as a function of \thbr,
for the case in which predicted masses are derived from the \mtx\
relation for the broadband spectral fit temperature.  Values of \thbr\
calculated both with cool lumps excised (squares) and without excision
(circles) appear in this figure.  Error bars on \thbr\ are omitted for
clarity.  While there is some correlation, such that clusters with
more temperature structure (larger \thbr) tend to be more massive than
predicted by the mean \mtx\ relation, the trend is weak and has
substantial scatter.  Excising the cool lumps weakens the trend
further.  Figure \ref{fig:f08} shows the same comparison for the
$RES_{cool}$ measure instead of \thbr, and in this case we again find
that excising the cool lumps has an even larger effect for \thbr\,
although in both cases accounting for temperature substructure does not
greatly reduce scatter in mass offset.  

\subsection{Effects of Masking Strategy}
\label{sec:masking}

\begin{figure}[p]
  \centering
  \includegraphics[width=0.9\textwidth, trim=0mm 0mm 0mm 0mm, clip=true]{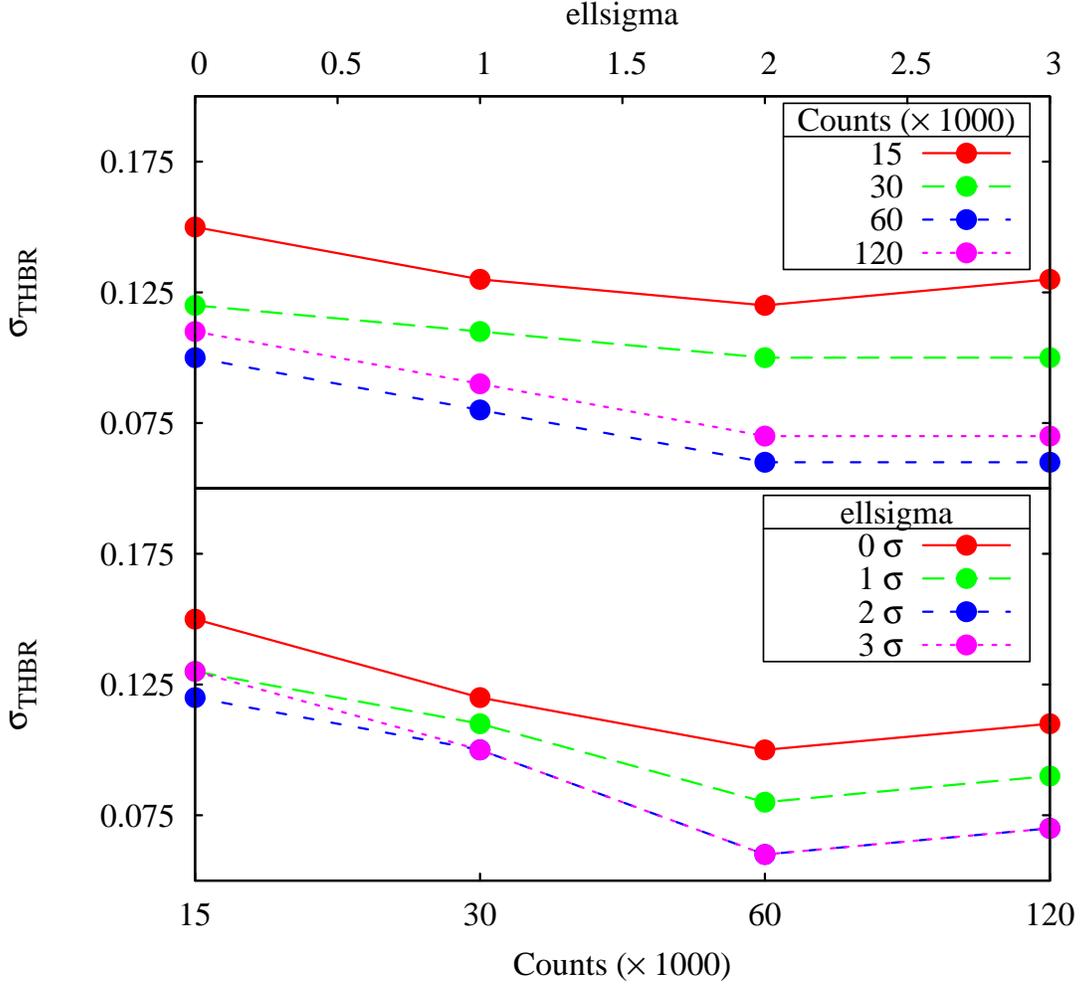}
  \caption{(\emph{top}) Decline in the \thbr\ standard deviation
    $\sigma_{\rm THBR}$ as wavdetect's ellsigma
    parameter ranges from 0 to 3.  These are plotted for a family of
    simulated observations of increasingly higher quality, from
    approximately 15k counts to approximately 120k counts.
    (\emph{bottom}) Decline in $\sigma_{\rm THBR}$ as simulated \xmas2
    observations go from lowest-quality (approximately 15k counts) to
    highest-quality (approximately 120k counts).  These are plotted
    for a family of simulated observations with wavdetect's
    ellsigma parameter ranges from 0 to 3.
    \ColorRef}
  \label{fig:f09}
\end{figure}

We now examine the decline in \thbr\ and its variance as cool lumps
are excised, beginning with Figure \ref{fig:f09}, which focuses on
\thbr.  Here, we plot the statistic's standard deviation $\sigma_{\rm
 THBR}$ as a function of the two dimensions along which we adjust our
analysis pipeline, with the top panel devoted to the masking strategy,
and the bottom panel devoted to the count level.  Focusing on the top
panel, we see that increasing the \verb|ellsigma| parameter of the
CIAO \verb|wavdetect| tool from 0 to 3 reduces $\sigma_{\rm THBR}$
from 0.15 down to 0.13, when the observations are
relatively ``poor'' (with $\sim$15K counts).  With higher-quality
observations of 60K or 120K counts, the decline is in
$\sigma_{\rm THBR}$ greater, going from 0.12 down to 0.07 as the
 \verb|ellsigma| parameter rises from 0 to 3, and $\sigma_{\rm THBR}$ 
for full masking ends up being significantly less than the value of 0.10 
observed in the \cite{cavagnolo_bandpass_2008} sample.

\begin{figure}[p]
  \centering
  \includegraphics[width=0.9\textwidth, trim=0mm 0mm 0mm 0mm, clip=true]{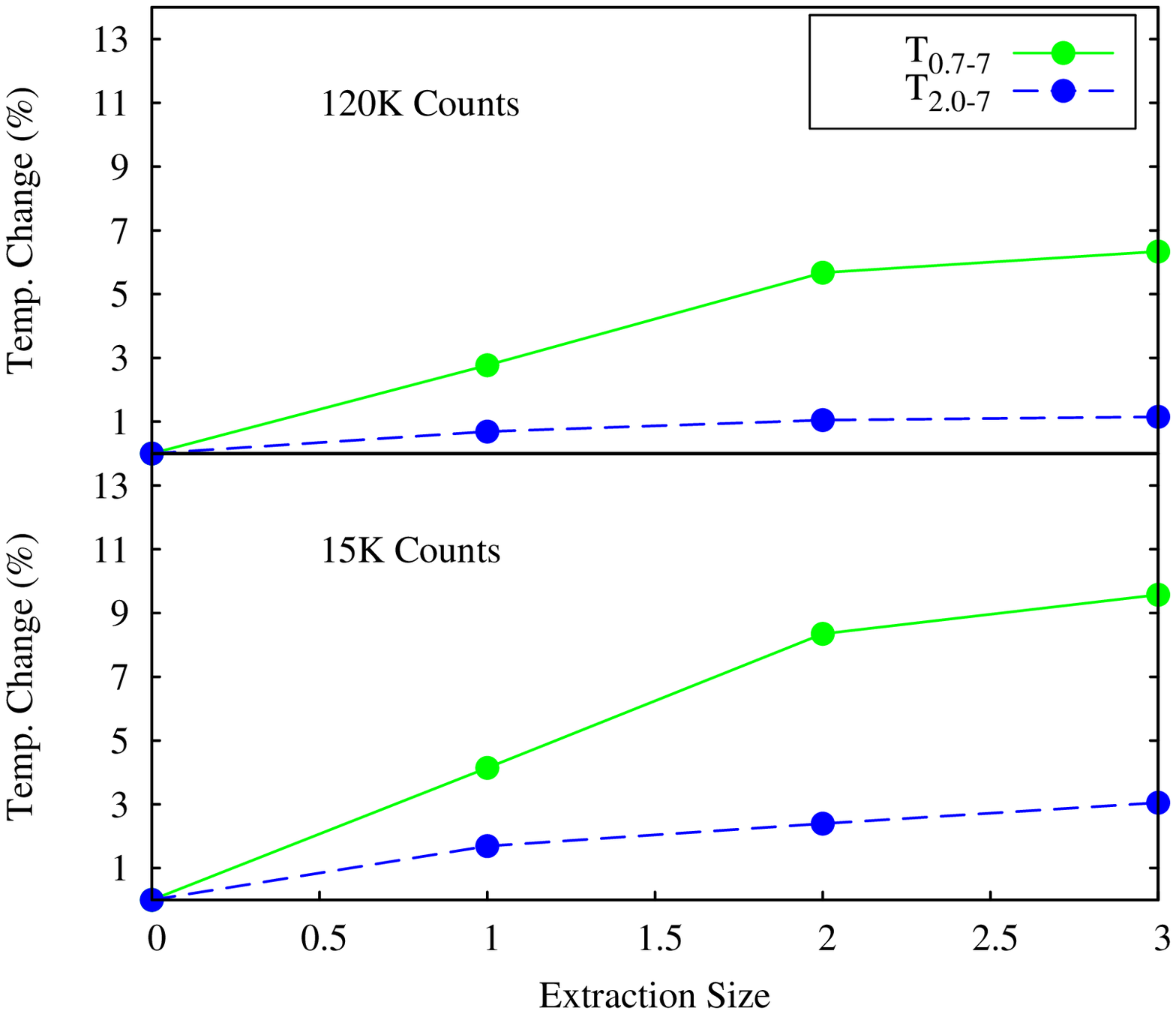}
  \caption{Relative change in spectral fit temperatures as the cool
    lumps are removed from the simulated clusters for high-quality
    simulated observations.  (\emph{top}) Observations with
    approximately 120K counts.  (\emph{bottom}) Observations with
    approximately 15K counts.  Note that these are for observations in
    an aperture corresponding to $R_{2500}$.\ColorRef}
  \label{fig:f10}
\end{figure}

Figure \ref{fig:f10} helps show why the temperature ratio \thbr\ and
its variance decline as the cool lumps are removed.  It plots the
sample average of the relative change in temperature as the CIAO
\verb|wavdetect| \verb|ellsigma| parameter is increased.  As more of
each cool lump is excised, both the broadband spectral fit
temperature, and the hardband fit temperature increase.  However, the
increase is significantly larger for the broadband temperature, as
the cool lumps' contribution to the flux is already largely excluded
from the hardband fits.  The top panel in this figure shows this
effect for spectra with approximately 120K counts, while the bottom
panel is for spectra with approximately 15K counts.

\begin{figure}[p]
  \centering
  \includegraphics[width=0.9\textwidth, trim=0mm 10mm 0mm 10mm, clip=true]{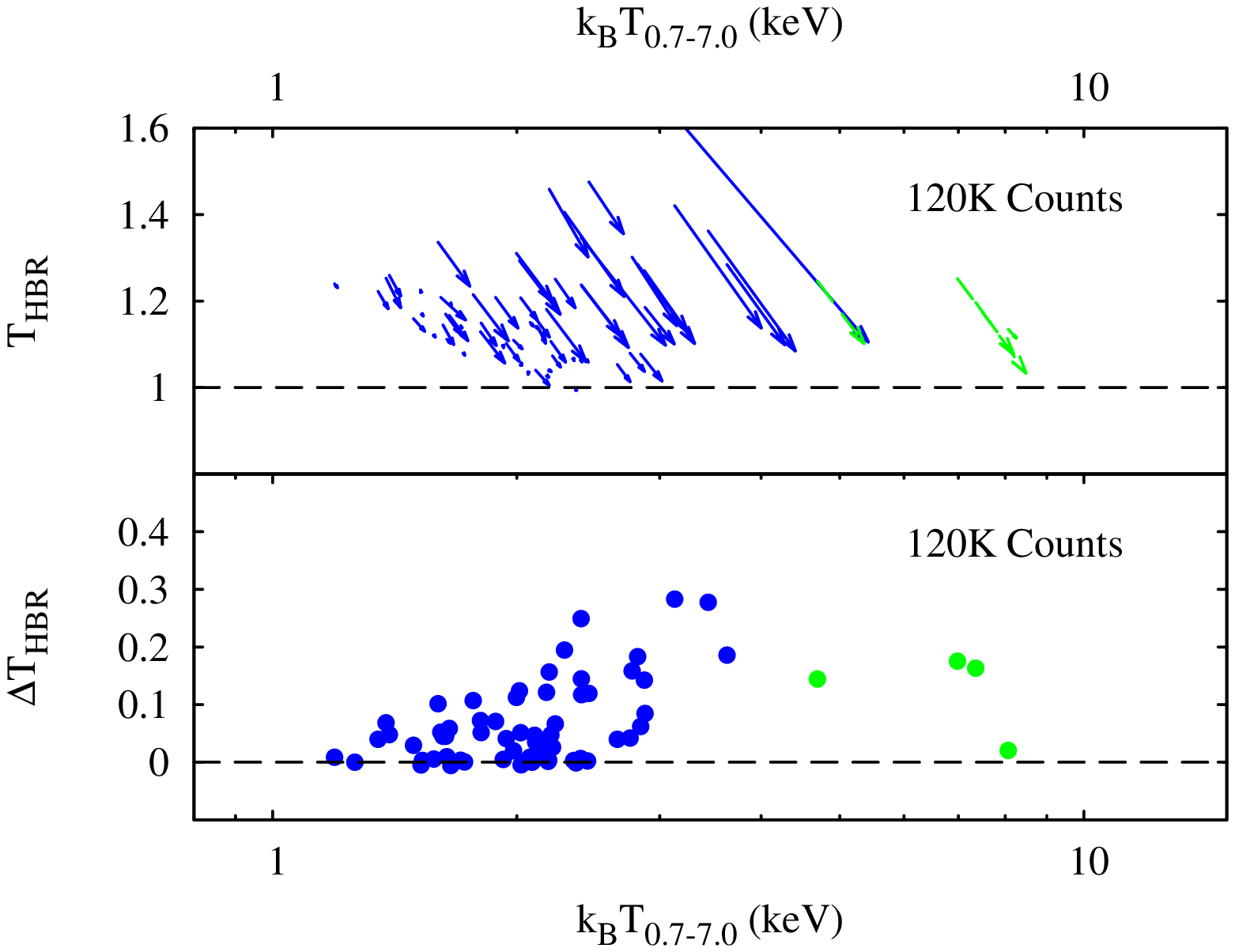}
  \caption{(\emph{Top}) Shift in the \tbroad-\thbr\ plane as
    wavdetect's ellsigma parameter ranges from 0 to
    3.  Clusters from the B04 sample are depicted with solid arrows,
    while clusters from the D09 sample are depicted with long-dashed
    arrows.  broadband spectral-fit temperatures are estimated within
    an aperture set at $R_{2500}$.  Notice that the distribution of
    most of the points shifts significantly to higher spectral fit
    temperatures and smaller temperature ratios as cool lumps are
    excised.  Notice also that some simulated clusters are unafflicted
    by cool lumps, so that their net shift is 0.  (\emph{Bottom}) The
    \thbr\ component (y-axis) of the shift presented in the top panel.
    Note that these are for simulated \xmas2 observations of maximum
    quality, having approximately 120K counts.  \ColorRef}
  \label{fig:f11}
\end{figure}

Another way of visualizing the effect of more aggressive masking is
depicted in Figure \ref{fig:f11}.  The top panel in this figure is
similar to Figure \ref{fig:f06} in that it occupies the
\tbroad--\thbr\ plane, with \thbr\ plotted as a function of \tbroad\
for our simulated clusters.  Arrows illustrate the shift in \thbr\ and
in \tbroad, as the cool lumps are excised.  Some of the clusters
experience very large shifts in both quantities, while others
experience no shifts at all.  The former are associated with clusters
that have many well-defined and bright cool lumps, while the latter
correspond to those clusters that are completely free of cool lumps.
The bottom panel in this figure shows just the size of this shift for
the \thbr\ statistic, from which we can see that some clusters indeed
have a shift of precisely 0.  Again, these are clusters whose
simulated X-ray observations are unchanged after applying the CIAO
\verb|wavdetect| tool because it finds no sources to mask out.

\subsection{Substructure Measures and \thbr}

Various researchers have established a clear correlation between
morphological measures of cluster dynamical state and $\delta \ln M$
\citep{jeltema_cluster_2008, ventimiglia_substructure_2008,
piffaretti_valdarnini_substructure_2008, yang_influence_2009},
a correlation which is less apparent in our study of spectral measures
of substructure.  Similiarly, \cite{cavagnolo_bandpass_2008} found 
a correlation between the temperature ratio and structure for their \chandra\ sample, in that merging events are associated with 
elevated \thbr.  To probe this issue further we focus on \thbr\ and
compare it to several metrics for cluster substructure.  These are the
centroid variation $w$, the axial ratio $\eta$, and the power ratios
$P_{20}$ and $P_{30}$ \citep[see][and references therein]{ventimiglia_substructure_2008}.  
The centroid variation $w$ measures the
skewness of a cluster's two-dimensional photon distribution by
calculating the variance in a series of isophotes for the cluster
surface-brightness map.  The axial ratio $\eta$ measures a cluster's
elongation, which tends to increase during merger events.  The power
ratios $P_{20}$ and $P_{30}$ decompose a cluster's surface-brightness
image into two-dimensional multipole expansions, capturing different
aspects of a cluster's geometry.  $P_{20}$ relates to the ellipticity
in an image and is similar to the axial ratio $\eta$, while $P_{30}$
measures the ``triangularity'' in an image.

\begin{figure}[p]
  \centering
  \includegraphics[width=0.9\textwidth, trim=0mm 10mm 0mm 10mm, clip=true]{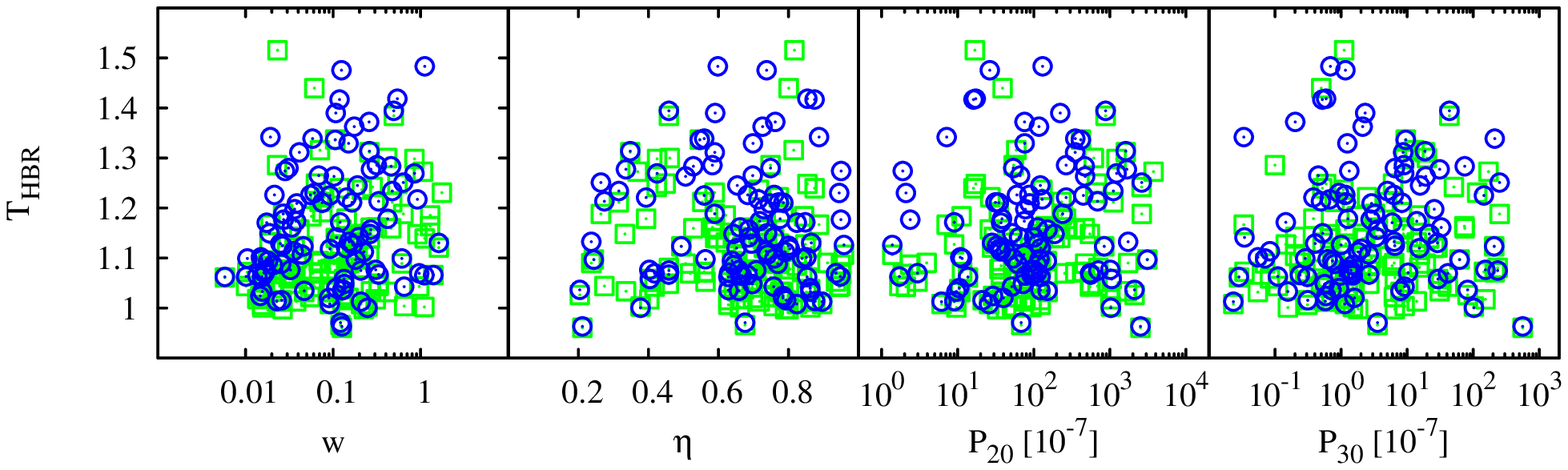}
  \caption{Relationship between \thbr\ and four measures of
    substructure: the axial ratio $\eta$, the centroid variation $w$,
    and the power ratios $P_{20}$ and $P_{20}$.  Squares correspond to
    simulated \xmas2 observations that are processed by wavdetect with
    ellsigma set to 3 (full masking), while circles correspond to
    observations that have ellsigma set to 0.  \ColorRef}
  \label{fig:f12}
\end{figure}

In \cite{ventimiglia_substructure_2008} we calculated these
morphological substructure metrics for a superset of the B04 sample of
simulated galaxy clusters and compared them to departures from the
\mtx\ relation (see Fig. 8 in that paper).  Here we use the same
substructure measures for the 114 B04 clusters for which we are able
to calculate \thbr\ and compare the results.  These are presented in
Figure \ref{fig:f12}.  Whereas in \cite{ventimiglia_substructure_2008}
there is a clear correlation between morphological substructure and
$\delta \ln M$, we find in this study that there is little or no
correlation between substructure and \thbr\, for our B04 sample.

\begin{figure}[p]
 \centering
 \includegraphics[width=0.9\textwidth, trim=0mm 0mm 0mm 0mm,
clip=true]{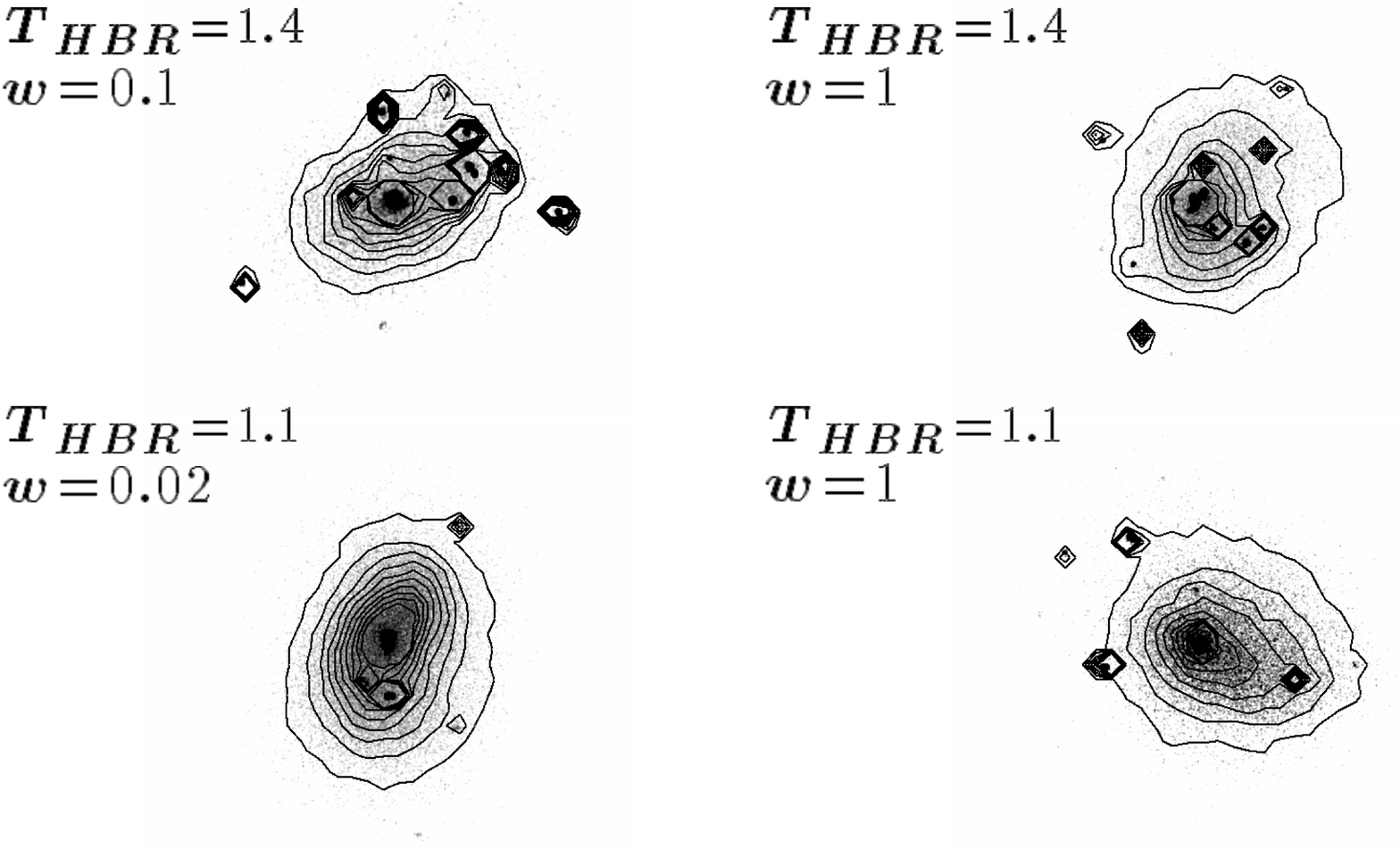}
 \caption{Surface-brightness contours for four clusters drawn from
   our simulation sample, illustrating that temperature ratio and 
   centroid shift are both imperfect measures of relaxation.  All four
   clusters are at nearly the same temperature, with $k_BT \simeq
   3.2$.  The two in the left column exhibit little centroid shift
   $w$, while the two in the right column have centroid shift
   near the maximum for the sample.  The two clusters in the bottom
   row have low \thbr\, while the two in the top row have higher
   \thbr\, suggesting the presence of multiple temperature
   components.}
 \label{fig:f13}
\end{figure}

In order to understand how temperature structure and morphological
structure can be uncorrelated, we looked at four simulated clusters having
approximately the same temperature ($k_BT \simeq 3.2 keV$) and
occupying the relative extremes in \thbr\ and in the centroid
variation $w$.  Two were selected for relatively low \thbr\
($\simeq1.1$) but large variation in $w$ ($\simeq 0.02$--$1.0$).
Two were selected for large \thbr\ ($\simeq1.4$) but again large
variation in the morphological structure parameter $\omega$ ($\simeq
0.1$--$1.0$).  These clusters are presented in Figure \ref{fig:f13}
with surface-brightness contours overplotted and with the associated
values for \thbr\ and $w$.  The two clusters in the left column appear
more symmetric in their contours, while the two in the right column
exhibit noticeable centroid shift.  However, the morphologically
apparent structure in the lower right cluster is not observed
spectrally in the temperature ratio \thbr.  Evidently, neither
morphology nor \thbr\ is a perfect measure of relaxation in our
simulation sample, as it contains clusters with small centroid shift
$\omega$ and obvious substructure, as in the upper left of this
figure.  And, it contains clusters with large $w$ that are nearly
isothermal, as in the lower right of this figure.

\section{Summary}

We used a sample of galaxy clusters simulated with radiative cooling
and supernova feedback, along with simulated \chandra\ X-ray
observations of these clusters, to study temperature inhomogeneity as
a signature of cluster dynamical state.  Specifically, we adopted two
methods of quantifying temperature inhomogeneity spectroscopically,
the temperature ratio \thbr\
\citep{mathiesen_four_2001,cavagnolo_bandpass_2008} and the cool residual
$RES_{cool}$.  The former is the ratio of a hardband X-ray
spectral-fit temperature to a broadband temperature, and becomes
greater than 1 for clusters whose ICM contains cool, over-luminous
sub-components.  The latter is the excess broadband count rate
relative to the count rate predicted by a model fit to the hard X-ray
band.  Though our simulated clusters are typically less massive and
have lower temperatures than the \chandra\ archive clusters in
\cite{cavagnolo_bandpass_2008}, we find that their temperature ratios
\thbr\ occupy generally the same distribution as the observed
clusters.

We also looked for an opportunity to combine \thbr\ and the cool
residual with the mean mass-temperature relation to obtain better mass
estimates than are achieved just with the scaling relation alone.  We
find, however, that while both \thbr\ and the $RES_{cool}$ are
correlated with offset from the \mtx\ relation, 
these correlations are weak, at least for this sample. We conclude that
these measures of temperature inhomogeneity are not very effective at
reducing scatter in the mass-temperature relation.

Finally, we note a particular difficulty that arises when trying to
use clusters from hydrodynamic simulations to calibrate
scatter-correction observables based on temperature inhomogeneity,
such as \thbr.  Historically, simulated clusters have tended to
exhibit their own kind of ``over-cooling problem'', in which dense
lumps of luminous, cool gas associated with merged sub-halos appear.
These cool lumps are often excised from simulated clusters before
their global properties are measured, but masking them appears to make
the remaining temperature structure of the simulated clusters overly
homogeneous.  This finding suggests that real clusters may have cooler
sub-components, that are more diffuse and less concentrated than in
their simulated counterparts.  Some physical process, perhaps thermal
conduction, 
turbulent heat transport \citep[e.g.,][]{Dennis_Chandran_2005, Parrish_Quataert_Sharma_2010, Ruszkowski_Oh_2011},
 or a more aggressive form of feedback, prevents cool lumps from forming in real clusters and might not completely eliminate those temperature inhomogeneities.  Newer simulations incorporate treatments of conduction and AGN feedback, as well as more accurate treatments of mixing,
and it will be interesting to revisit these temperature inhomogeneity measures in simulated clusters when large samples of such simulated clusters become available.

\acknowledgments The authors wish to thank Stefano Borgani for
contributing the simulation data on which this project was based.
This work was supported by NASA through grants NNG04GI89G and
NNG05GD82G, through Chandra theory grant TM8-9010X, and through
Chandra archive grant SAOAR5-6016X.  Some of it was done at the Kavli Institute for Theoretical Physics in Santa Barbara, supported in part by the National Science Foundation under Grant No. NSF PHY05-51164.


\begin{deluxetable}{rccccc}
\tablecaption{\thbr\ \& $\sigma_{T_{HBR}}$ \label{tbl:01}}
\tablehead{counts & ellsigma & $\overline{T_{\rm HBR}}$\tablenotemark{a} &
 $\overline{\sigma}$\tablenotemark{a} & $\overline{T_{\rm
HBR}}$\tablenotemark{b} & $\overline{\sigma}$\tablenotemark{b}}
\tablewidth{0pt}
\tabletypesize{\small}
\startdata
15000  & 0 & 1.19 & 0.15 & 1.20 & 0.15 \\
15000  & 1 & 1.17 & 0.13 & 1.18 & 0.13 \\
15000  & 2 & 1.13 & 0.12 & 1.13 & 0.11 \\
15000  & 3 & 1.13 & 0.13 & 1.13 & 0.12 \\
30000  & 0 & 1.16 & 0.12 & 1.16 & 0.12 \\
30000  & 1 & 1.13 & 0.11 & 1.13 & 0.11 \\
30000  & 2 & 1.10 & 0.10 & 1.10 & 0.10 \\
30000  & 3 & 1.10 & 0.10 & 1.09 & 0.10 \\
60000  & 0 & 1.13 & 0.10 & 1.14 & 0.10 \\
60000  & 1 & 1.11 & 0.08 & 1.11 & 0.08 \\
60000  & 2 & 1.08 & 0.06 & 1.08 & 0.06 \\
60000  & 3 & 1.07 & 0.06 & 1.07 & 0.06 \\
120000 & 0 & 1.12 & 0.11 & 1.12 & 0.12 \\
120000 & 1 & 1.10 & 0.09 & 1.10 & 0.09 \\
120000 & 2 & 1.07 & 0.07 & 1.07 & 0.07 \\
120000 & 3 & 1.07 & 0.07 & 1.07 & 0.06 \\
\enddata
\tablenotetext{a}{All Clusters}
\tablenotetext{b}{Clusters with $k_BT_{\rm 2.0-7} > 2$ keV}
\end{deluxetable}



\end{document}